\documentclass[pre]{revtex4-2}
\usepackage{amssymb}
\usepackage{amsmath}
\usepackage{graphicx}
\usepackage{hyperref}
\usepackage{bm}
\ifx\affiliation\undefined 
\def\affiliation#1{\date{\normalsize #1}}
\usepackage[margin=1in]{geometry}
\usepackage[sort&compress,numbers]{natbib}
\fi
\def\energy{{\cal E}}

\begin{document}

\title{Asymmetric One-Dimensional Slow Electron Holes}
\author{I H Hutchinson}
\affiliation{Plasma Science and Fusion Center,\\ Massachusetts Institute of
  Technology,\\ Cambridge, Massachusetts, 02139, USA}

\ifx\altaffiliation\undefined\maketitle\fi 
\begin{abstract}
  Slow solitary positive-potential peaks sustained by trapped electron
  deficit in a plasma with asymmetric ion velocity distributions are
  in principle asymmetric, involving a potential change across the
  hole.  It is shown theoretically how to construct such asymmetric
  electron holes, thus providing fully consistent solutions of the
  one-dimensional Vlasov-Poisson equation for a wide variety of
  prescribed background ion velocity distributions. Because of ion
  reflection forces experienced by the hole, there is generally only
  one discrete slow hole velocity that is in equilibrium. Moreover the
  equilibrium is unstable unless there is a local minimum in the ion
  velocity distribution, in which the hole velocity then resides. For
  stable equilibria with Maxwellian electrons, the potential drop
  across the hole is shown to be
  $\Delta\phi\simeq{2\over 9}f'''{T_e\over e}({e\psi\over m_i})^2$, where
  $\psi$ is the hole peak potential, $f'''$ is the third derivative of
  the background ion velocity distribution function at the hole
  velocity, and $T_e$ the electron temperature. Potential asymmetry is
  small for holes of the amplitudes usually observed,
  $\psi\lesssim 0.5T_e/e$.
\end{abstract}
\ifx\altaffiliation\undefined\else\maketitle\fi  

\section{Introduction}

A Bernstein, Greene, Kruskal\cite{Bernstein1957} (BGK) mode is a
one-dimensional potential structure in a collisionless plasma that in
the mode's frame of reference is a steady nonlinear solution of the
Vlasov-Poisson system of equations relating electron and ion velocity
distribution functions, $f_e(v)$, $f_i(v)$, to the electric potential
$\phi$. Electron holes are a subset of these BGK modes for which a
positive potential peak is sustained by a deficit of electrons trapped
by the potential\cite{Schamel1986,Hutchinson2017}, hence the
name. Normally electron holes are considered to be solitary waves in
which a single potential peak is embedded in a plasma that is uniform
far from the peak. When there is negligible reflection of ions by the
potential, for example because the ions' mean velocity (in the hole
frame) far exceeds their distribution width, electron holes are
symmetric about the potential peak. This symmetry is required by the
fact that the trapped-electron distribution must be symmetric in
velocity, and the passing-electron and ion densities are functions
only of potential, regardless of any velocity distribution
asymmetry. Such holes can move at essentially any velocity relative to
the ions greater than a few ion sound speeds, up to the electron
thermal speed. A considerable theoretical literature on symmetric
electron holes has established many of their important
properties (e.g.\cite{Eliasson2006,Krasovsky2003,Dupree1982,Turikov1984,Chen2004,Ng2006,Hutchinson2018a,Hutchinson2019,Zhou2018,Hutchinson2021b}). Moreover,
space plasma observations of sufficient time resolution now often
observe fast-moving potential peaks interpreted as electron holes (e.g.\cite{Ergun1998,Ergun1999,Malaspina2013,Malaspina2014,Malaspina2018,Malaspina2019,Steinvall2019,Graham2016,Tong2018,Lotekar2020}).

By contrast, when there is reflection of ions, because $f_i(v)$ is
non-negligible near $v=0$ (which we call a `slow' electron hole
situation) it can produce a net interaction force exerted by the
potential hill on the ions, $F_i$, when the ion distribution is
asymmetric in velocity. Since the potential is sustained in place only
by the plasma particles, in equilibrium the ion force must be balanced
by an equal and opposite force exerted by the potential on the
electrons, $F_e$, making the total zero: $F_i+F_e=0$. The electron
force $F_e$ can be non-zero in equilibrium only if there is some
(positional) asymmetry in the potential. In fact, as we shall show,
for given $f_i$ and $f_e$ in some other fixed frame (e.g.\ the `ion'
frame in which mean ion velocity is zero), there is generally only one
discrete mode velocity that gives rise to an equilibrium $F_i+F_e=0$.

Moreover, even if $F_e+F_i=0$, for example when there is a velocity
about which both $f_e$ and $f_i$ are symmetric, the equilibrium it
represents may be \emph{unstable}. It has been
established\cite{Hutchinson2021c} that electron holes interacting with
single-humped ion distributions are essentially always
unstable\cite{Eliasson2004}, accelerating the hole velocity till ion
reflection becomes negligible\cite{Zhou2016}, or until the hole itself
is trapped by coupling to an ion acoustic
soliton\cite{Saeki1998,Zhou2017}. It is crucial for the long term
persistence of an electron hole experiencing ion reflection, that it
be stable against such self-acceleration. A number of recent
spacecraft plasma observations have reported slow holes for which ion
reflection should be important\cite{Graham2016,Steinvall2019,Lotekar2020,Kamaletdinov2021}.

Ion reflection dictates the equilibrium slow electron hole velocity,
and asymmetric ion velocity distributions make electron hole
potentials asymmetric. In particular, the ion density will generally
be different on either side of the potential peak, requiring the
electron density there likewise to be asymmetric so as to satisfy
quasi-neutrality far from the hole. This will generally require there
to be a potential difference $\Delta \phi=\phi(+\infty)-\phi(-\infty)$
across the hole that persists into the quasineutral region.

\begin{figure}[htp]
  \centering
  \includegraphics[width=0.6\hsize]{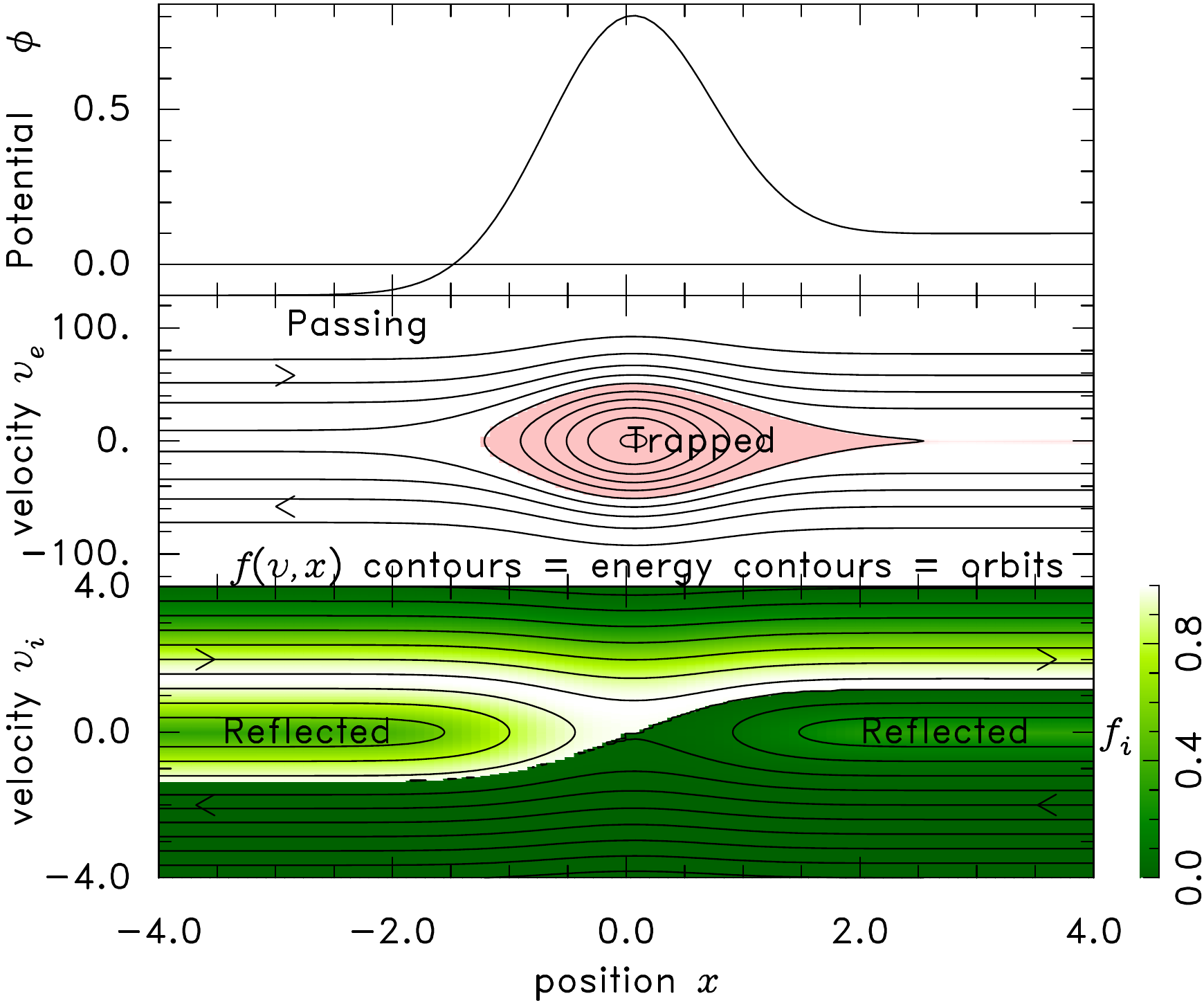}
  \caption{Schematic of a hypothetical asymmetric electron hole.}
  \label{fig:electronholeorbits}
\end{figure}
Fig.\ \ref{fig:electronholeorbits} illustrates schematically the
potential, and electron and ion distribution function contours in
their respective phase-spaces. There is a region of closed (trapped) electron
orbits whose distribution function $f_e(x,v_e)$ is set by the
formation conditions of the structure. It is generally lower than the
nearby untrapped (passing) $f_e$, which is set by boundary conditions and the
constancy of $f$ on orbits, because of the Vlasov equation. The
trapped electron deficit causes $n_e(x)<n_i(x)$ near $x=0$ and thereby
sustains the potential $\phi$. The ion distribution is everywhere set
by its value at the $x$-boundaries (infinity). In this illustration it
is a Maxwellian shifted by 1.5 velocity units. The outgoing
distribution is complicated by reflection and has discontinuities at
the transition between reflected and unreflected orbits.

The problem addressed in the present work is this. Given prescribed
electron and ion velocity distributions incoming at the boundaries,
far from the potential structure, find a fully self-consistent
electron hole equilibrium (with a local potential peak), and show how
to calculate the electron hole velocity relationship to the
distributions, the potential drop $\Delta\phi$, and the relationship
between the hole potential and the trapped electron distribution
function. Furthermore, establish the circumstances under which such an
asymmetric equilibrium is stable.

In a recent publication\cite{Hutchinson2021c}, the equilibrium and
stability of slow electron holes with symmetric and asymmetric ion
distributions, and ion reflection, was analysed under the rather ad
hoc assumption that the hole potential is symmetric. It was found that
an essential ingredient of stability under this approximate ansatz was
that $f_i(v)$ should be double humped. And this is in accord with
recent space plasma observations\cite{Kamaletdinov2021}. The present
purpose is to proceed instead \emph{without} assuming the potential to
be symmetric, and thereby to complete and validate the analysis
of slow asymmetric electron holes. The findings substantially confirm
the prior simplified analysis.

\section{Theory Background}

In this paper the Vlasov equation will not actually be written
down. Instead its property that the distribution function $f(v)$ is
constant along orbits will be used directly. In a steady potential,
the particle energy is also a constant of the motion, and so $f$ is a
function of energy. Together with the knowledge that particle density
is $n=\int f dv$ these facts are sufficient to analyze equilibria.

Bernstein, Greene, and Kruskal, in their original
paper\cite{Bernstein1957} showed that one can formally solve to find
the required distribution functions to create \emph{any} arbitrary
mode potential shape $\phi(x)$, with an arbitrary number of minima and
maxima, as follows. Dividing the spatial domain into segments between
adjacent local minima and maxima ($\phi_{min}$ and $\phi_{max})$,
consider the ions in a segment of \emph{increasing} $\phi(x)$ and
suppose their velocity distribution to be known for all relevant
energy $\energy_i\equiv {1\over 2}m_iv^2+e\phi >e\phi_{min}$ and the
passing electron distribution to be known for
$\energy_e\equiv{1\over 2}m_ev^2-e\phi >-e\phi_{min}$. The electrons
reflected from this potential segment
$-e\phi_{max}<\energy_e<-e\phi_{min}$, entering from the right, have a
velocity distribution $f_{er}$ symmetric in $v$ and a function only of
energy. Their density must satisfy Poisson's equation
$n_{er}(\phi)+n_{ep}(\phi)-n_{i}(\phi)={\epsilon_0\over
  e}{d^2\phi\over dx^2}$, where subscripts $r$ and $p$ refer to
reflected and passing (unreflected) particles. Since
$n_{er}=2\int_0^{v_s} f_e(v)dv=
(2/m_e)\int_{-e\phi_{max}}^{-e\phi_{min}}f_e(\energy)/\sqrt{2(\energy+e\phi)}\,
d\energy$, and $n_{ep}(\phi)$, $n_{i}(\phi)$, and
${d^2\phi\over dx^2}$ are known, giving $n_{er}(\phi)$, an integral
equation governs the reflected part of $f_{e}(\energy)$ and can be
solved to find the unique required trapped-electron distribution
consistent with the specified potential profile. For the next segment
to the right, which has \emph{decreasing} $\phi(x)$, the roles of
electrons and ions are reversed, the entire $f_e$ is known and the
passing ion $f_{ip}$. One can thus find the required reflected ion
distribution from an integral equation. By this sequential process one
can in principle find the sequence of reflected (and trapped)
distribution functions that self-consistently satisfy Poisson's
equation and dependence of $f(v)$ only on energy, i.e.\ the steady
Vlasov equation.

Concerning a \emph{solitary} potential structure like an electron
hole, if the asymmetry $\Delta \phi$ becomes so great that it removes
the local potential maximum, giving rise to a monotonic potential
$\phi(x)$, and removing all local electron trapping\footnote{In this
  work we will distinguish between trapping and reflection. A
  reflected particle escapes to large distance; a trapped particle
  does not, being confined by a local potential energy minimum.}, then
the structure is called a Double-Layer.  Double-layers have a long
history of study since their first experimental observation and
analysis by Langmuir\cite{Langmuir1929}. Although the possibility of
asymmetric solitons, with local potential minima or maxima has been
noted in these and other double layer studies\cite{Schamel1986},
almost all of the analysis assumes that the double-layer potential is
monotonic or occasionally has a local minimum (i.e.\ an ion hole, or
an electron-acoustic soliton\cite{Vasko2017a}).  A (monotonic)
double-layer has a single potential segment, and the approach of BGK
described in the previous paragraph describes how, given $\phi(x)$ and
the entire incoming distribution of the reflected species on one side,
the required distribution of the other species on the other side can
be found. Variations around the BGK integral equation method appeared
in the early development of double-layer analysis
\cite{Montgomery1969,Knorr1974}. They were joined by approaches that
express the shape of the velocity distribution in terms of a few
fluid-like parameters such as reflected species effective temperature
and passing mean velocity. These often used what is essentially BGK's
differential equation method and the requirement of net charge and
force neutrality in the form of boundary conditions for given
potential drop $\Delta \phi$ (e.g.\cite{Schamel1983}). The influential
model of Perkins and Sun\cite{Perkins1981}, for example, showed that
provided the passing particle distributions are chosen appropriately,
no net electric current need flow across the double-layer, which had
previously been in doubt; but their model had only one adjustable
parameter governing the reflected ions, thereby constraining both
$\Delta \phi$ and the trapped ion parameter to be unique functions of
the passing electron to ion temperature ratio. One should beware of so
called nonlinear dispersion relations like this; they arise because of
artificially prescribing the shape of the trapped distribution.
Double-layer analysis is well summarized in the extensive reviews of
Raadu\cite{Raadu1989,Raadu1988} and their references.

An electron hole, though, such as illustrated in Fig.\
\ref{fig:electronholeorbits}, has two segments and a single local
potential maximum, thus occasionally being referred to as a
``Triple-Layer''. Moreover, neither the double-layer analyses nor the
BGK sequential integral equation approach show how to deal with a
situation in which the incoming velocity distributions of the
particles on either side of the potential structure are broad but
known, and we wish to solve instead for the potential $\phi(x)$ when
it is unknown. This is nearest to the situation encountered in space
observations, on which most of the electron hole experimental research
is currently focussed, and in which satellites generally measure the
ion and electron distribution functions in the background plasma. It
is also what is needed to initialize a consistent slow electron hole
in a simulation with prescribed particle velocity distributions. And
it is the subject of the present work. Our interest includes the
stability of the electron hole velocity, which is vital in this
context for a slow electron hole to persist. All of the considerations
here are purely one-dimensional.

\section{Problem specification and  approach}

\subsection{Specifying the ion distribution}
We begin by supposing that the incoming ion velocity distribution far
from the hole is known and the potential is
steady in the rest frame of the hole.  The distribution at arbitrary
position $x$ is then governed by
$f_i(x,v)=f_{i}(\infty,v_\infty)$, with energy conserved along orbits
giving total ion energy $\energy=mv^2/2+e\phi=mv_{\infty}^2/2+\phi_\infty$, and
$v_{\infty}$ and $\phi_\infty$ corresponding to whichever side of the
hole the ion entered.  Denote the sign of $x$ (the position relative
to the potential peak at $x=0$) by
$\sigma_x(=\pm1)$. At $x$, all inward moving ions entered from the
same side $\sigma_{\infty}=\sigma_x$; but outgoing ions entered from
the other side $\sigma_\infty=-\sigma_x$ if they are passing, or the
same side $\sigma_x$ if they have been reflected. The sign of the
entering \emph{velocity} ($v_\infty$) is of course $-\sigma_\infty$.
So
$v_\infty=-\sigma_\infty\sqrt{2(\energy-e\phi_\infty)/m}=-\sigma_\infty\sqrt{v^2+2e(\phi-\phi_\infty)/m}$,
and consequently
\begin{equation}\label{ni}
  \begin{split}
  n_i(x)=\int f_i(x,v) dv=\int
  f_{i}(\sigma_\infty\infty,-\sigma_\infty\sqrt{v^2+2e(\phi-\phi_\infty)/m})\,dv.
    \end{split}
\end{equation}
Evaluation of this integral requires knowledge of the peak potential
height $\psi$ (at $x=0$) because outgoing ions of energy
$\energy<e\psi$ have been reflected, while those with $\energy>e\psi$ have
not. Thus $\sigma_\infty$ changes sign at $\energy=e\psi$. This change
generally causes a discontinuity in
$f_{i}(x,v)$.
\begin{figure}[htp]
  \centering
  \includegraphics[width=.6\hsize]{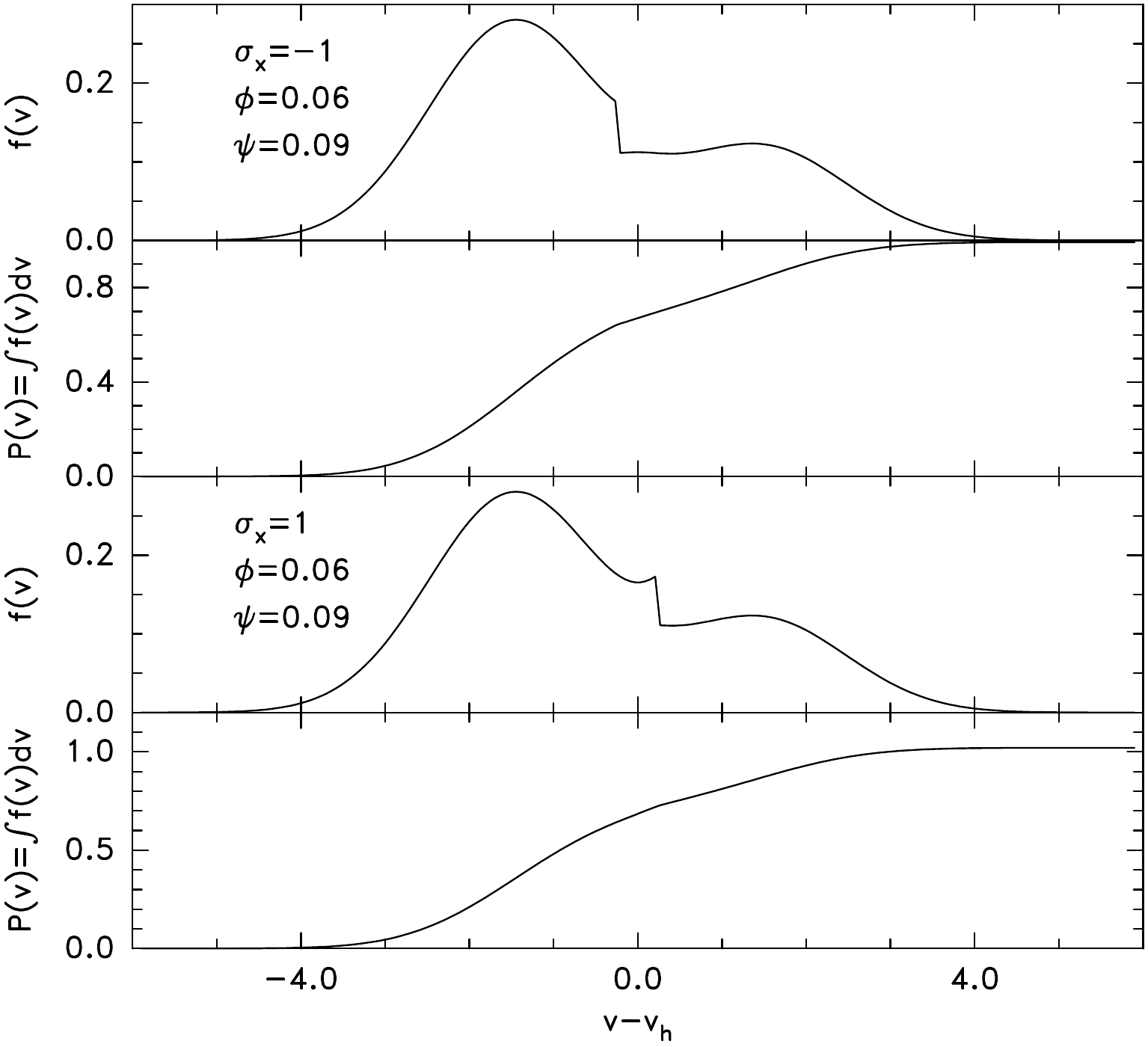}
  \caption{Illustration of ion distributions on either side of the
    potential hill, and their integral, which is the cumulative
    probability distribution $P(v)$, and equals $n_i(\phi)$ at $v=+\infty$.}
  \label{fig:fiplots}
\end{figure}
Fig. \ref{fig:fiplots} illustrates these features for an incoming ion
velocity distribution consisting of the sum of two Maxwellians having density,
mean velocity, and temperature respectively: $(0.3,1.5,1)$ and
$(0.7,-1.5,1)$, and potential peak $\psi=0.09$. In the codes and
this and all other plots we use units normalized to Debye length
$\sqrt{\epsilon_0T_0/n_ie^2}$, reference ion thermal energy $T_0$ and
thermal velocity $\sqrt{T_0/m_i}$, and densities are unity when
$\phi=\psi=0$. In these units, the ion mass is 1 and the ion charge is
$e=1$, and potential is in units $T_0/e$. The electron temperature
$T_e$ is equal to $T_0$ unless otherwise noted. Recognize that $\phi_\infty$ is
initially unknown, and different for different sides $\sigma_\infty$.

In view of the energy conservation that yields eq.\ (\ref{ni}), it seems
best to regard $f_{i}(\infty)$ as a fixed function of energy $\energy$
and velocity sign, regardless of $\phi_\infty$, so as to
make the passing ion distribution independent of
$\Delta\phi=\phi(+\infty)-\phi(-\infty)$. To do
so requires us to prescribe $f_i(\infty)$ at some negative values of
energy, since for non-zero $\Delta\phi$, when
$\bar\phi_{i\infty}=(\phi(+\infty)+\phi(-\infty))/2$ is taken to be
the zero of potential, the lower side's potential becomes negative and
we need the incoming ion distribution there down to zero velocity. We
therefore regard a function $f_{i\infty}(v)$ to be prescribed, and take
\begin{equation} 
  \label{eq:fiinf}
  f_i(\sigma_\infty\infty,\energy)=
  \begin{array}{ll}f_{i\infty}(-\sigma_\infty\sqrt{2\energy/m})
  & \mbox{for}\quad \energy\ge 0\\
  f_{i\infty}(0) & \mbox{for}\quad \energy< 0.
  \end{array}
\end{equation}
Then we will later display the chosen distant ion distributions by a plot of
$f_{i\infty}(v)$.

\subsection{Determining distant potential asymmetry}
\label{phisymmetry}
\begin{figure}[htp]
  \centering
  \includegraphics[width=0.6\hsize]{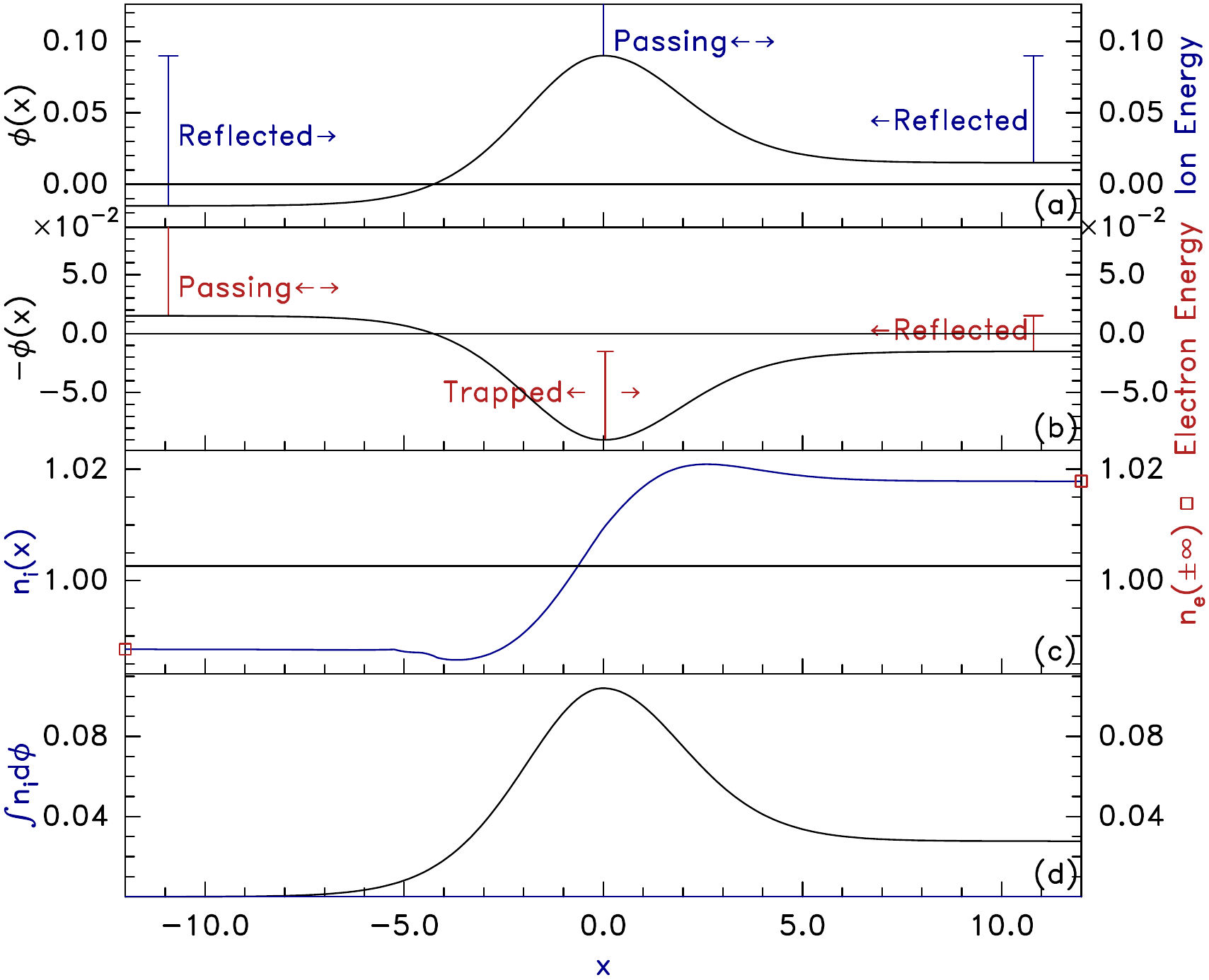}
  \caption{Illustrative asymmetric electron hole parameters (a)
    $\phi(x)$, which is ion potential energy; (b)
    $-\phi(x)$, electron potential energy; (c)
    $n_i(x)$ ion density; (d) `Classical' potential giving force.
  The vertical bars in (a) and (b) indicate ranges of particle energy
  that are passing, reflected, or trapped. The two points in (c) show
  the matching distant electron density. The nonzero value of
  $\int_{-\infty}^{+\infty} n_id\phi$ shows this hole is subject to
  non-zero ion force.}
  \label{fig:explot}
\end{figure}

Knowledge of the incoming
$f_{i}(\sigma_\infty\infty,-\sigma_\infty|v_\infty|)$ and $\psi$ is
sufficient to determine also the distant \emph{outgoing} distribution
as a function of energy, when both of $\phi(\pm\infty)$ are known.
The ion densities $n_{i\infty}(\sigma_{\infty}\infty,\psi)$ on either
side then generally differ, and depend on $\psi$. Figure
\ref{fig:explot} illustrates the result for the same incoming ion
distribution and potential peak height as Fig.\ \ref{fig:fiplots}.

For a solitary structure like an electron hole, the plasma must be
neutral ($n_{i\infty}-n_{e\infty}=0$) at distant positions
on both sides, to bring the external potential curvature to
zero. Consequently, if the electron distribution $f_e(\energy_e)$ is
known, and therefore $n_e(\phi)$ is a known fixed function, then the
two simultaneous neutrality requirements are sufficient in principle
to determine the two $\phi_\infty$ values. In practice it is
convenient to choose the electron parameters so that $n_e(\phi)$ is an
easily invertible function. A natural choice is to assume it has
Boltzmann dependence $n_e\propto \exp(e\phi/T_e)$, so
\begin{equation}\label{fedist}
\Delta \phi=\phi_{i}(+\infty)-\phi_i(-\infty)=
(T_e/e)\ln(n_i(+\infty)/n_i(-\infty)).  
\end{equation}
But for numerical solution one could make other choices. Indeed, one
could consider electron distributions that are asymmetric in incoming
velocity, so that when $\Delta\phi$ is non-zero, giving rise to
electron reflection, the electron density then depends on $\sigma_x$ (as
well as $\phi$). A treatment that performed the integration over
specified \emph{electron} distribution, like eq.\ (\ref{ni}), would
then be required, imposing moderate extra computational effort.

In any case, since the distant ion density $n_i(\pm\infty)$ is itself
a nonlinear function of $\phi(\pm\infty)$, $\Delta\phi$ solutions in
general have to be found by iteration. In the present work it is
assumed that $n_e=n_{e0}\exp(e\phi/T_e)$. That is a good approximation
for Maxwellian electrons and weak current density. We shall also take
$\psi$ to be fixed relative to the mean
$\bar\phi_{i\infty}=[\phi(+\infty)+\phi(-\infty)]/2$, which is taken
to be the zero of potential. The difference $\Delta\phi$ evolves
during solving iterations. Newton's method applied to the residual
$(T_e/e)\ln(n_i(+\infty)/n_i(-\infty))-\Delta\phi$ is observed to
converge to numerical integration accuracy in fewer than
10 iterations. Once $\Delta\phi$ is converged, $n_{e0}$ is determined
by $n_{e0}=n_i(\pm\infty)/\exp(\pm e\Delta\phi/2T_e)$.  This procedure
produces potential limits that are consistent with the incoming ion
and electron distributions and $\psi$. Fig.\ \ref{fig:explot}
illustrates the spatial dependencies using the converged
$\Delta\phi$, when the hole velocity $v_h$ is zero in the ion frame.

\subsection{Poisson's equation and force balance}

A full solution of a one-dimensional electron hole shape satisfies
Poisson's equation
\begin{equation}
  \label{eq:poisson}
  \epsilon_0{d^2\phi\over dx^2} = -\rho,
\end{equation}
where $\rho$ is the charge density $\rho=e(n_i-n_e)$. When $\rho$ is a
function only of potential not directly $x$, as is the case here, such
a differential equation can be integrated once as
$[{\epsilon_0\over 2}({d\phi\over
  dx})^2]_{\phi_0}^\phi=-\int_{\phi_0}^\phi\rho d\phi\equiv-V$; and
the second integral
$[x]_{\phi_0}^\phi=\pm\int_{\phi_0}^\phi \sqrt{\epsilon_0/2|V|}d\phi$
provides the solution in the form $x(\phi)$. In the soliton context,
$V$ is often called the `Classical' or `Sagdeev' potential. The key
boundary conditions of a solitary solution are that $V$ be zero at the
extrema of $\phi$, including at $x\to\pm \infty$ where
$dV/d\phi=e(n_i-n_e)=0$ (just discussed) and $d^2V/d\phi^2\le 0$ (to
ensure the nearby $V$ is non-positive). Behind the mathematics, though
\cite{Andrews1971a}, $V=\int\rho{d\phi\over dx}dx$ is
physically minus the integrated force exerted on the charge by the
electric field; and ${\epsilon_0\over 2}({d\phi\over dx})^2$ is the
Maxwell stress, whose difference expresses the same quantity
$V = -\int\epsilon_0{d^2\phi\over dx^2}{d\phi\over
  dx}dx=-\left[{\epsilon_0\over2}\left(d\phi\over dx\right)^2\right]
$. Fig. \ref{fig:explot}(d) shows the integration $\int n_id\phi$ for
ions alone, representing minus the force on the ions.  For a
\emph{symmetric} electron hole (or other soliton) the forces on the
two sides cancel by symmetry; but an \emph{asymmetric} potential has
no guaranteed force cancellation. And in fact the total force (per
unit transverse area) $F$ will generally be non-zero, causing hole
acceleration, except when the hole has a particular velocity relative
to the specified incoming distributions.  There is thus an additional
criterion for a steady equilibrium that enforces a particular hole
velocity so as to satisfy force balance. A symmetric-potential hole
satisfies this criterion when it has zero or symmetric reflected
particle velocity distribution in the hole frame.
Fig. \ref{fig:explot} in fact has non-zero $F$ and so does \emph{not
  satisfy} force balance. It is not actually an
\emph{equilibrium}. The potential structure would be subject to
acceleration.

Satisfying $F=0$ places no direct constraints on the \emph{trapped}
electron distribution, having energy
$-e\psi\le \energy_e \le -e|\Delta \phi/2|$. The reason is that
trapping enforces symmetry of the distribution, so the trapped
densities $n_{et}$ (and electron charge-densities) on the two sides of
the potential peak at the same potential are equal, and the two contributions
$-e\int n_{et} {d\phi\over dx} dx$ are equal and opposite (as are the
passing electron contributions). The remaining
contribution of electrons to $F$ arises from the integral of the
electron density over $\Delta\phi$, that is
\begin{equation}\label{Fe}
F_e=-V_e=e\int_{-\Delta\phi/2}^{\Delta\phi/2} n_{e}
d\phi=n_{e0}T_e[\exp(e\Delta\phi/2T_e)-\exp(-e\Delta\phi/2T_e)].
\end{equation} It
is the force of electron reflection from the potential
difference across the hole, and for the present Maxwellian electrons
is manifestly the electron pressure-difference across the hole.

The ion force also arises from reflection, in its case from either
side of the potential hill, and since there are no trapped ions, it
can be written simply
\begin{equation}\label{Fi}
F_i=-\sum_{\sigma_x=\pm1} \sigma_x e \int_{\phi(\sigma_x\infty)}^\psi
n_i d\phi,
\end{equation}
with $n_i$ given by eq.\ (\ref{ni}). If we regard the
electron and ion distant distributions as given in the fixed ion frame,
the only freedom we have to satisfy $F_e+F_i=0$ is to suppose that the
hole moves with some velocity $v_h$ relative to that frame, and that
$v_h$ is to be adjusted to satisfy force balance. This viewpoint is
intuitive, since the result of a non-zero total force will in fact be
hole acceleration, that is modification of $v_h$.

Thus, we must (again iteratively) search for a $v_h$ that gives $F=0$,
when $\Delta\phi$ is given as in section \ref{phisymmetry} by the
requirements on $n(\pm\infty)$. The result will be to find $v_h$,
$\Delta\phi$, that satisfy all the boundary conditions (including
force balance), without any constraints (beyond symmetry) so far on
the trapped electron distribution.

\begin{figure}[htp]
  \centering
  \includegraphics[width=0.6\hsize]{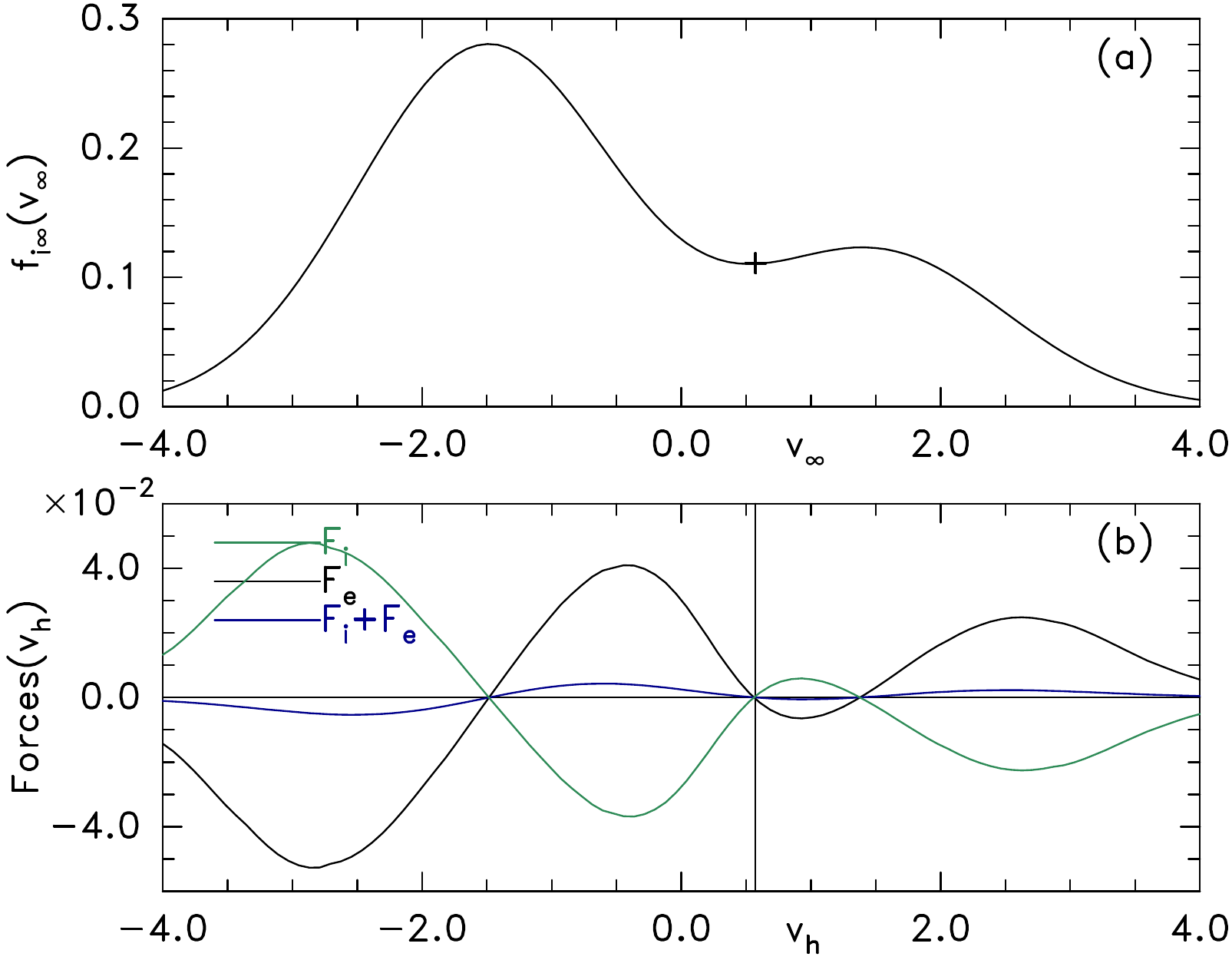}
  \caption{Distant velocity distribution of the ions ($f_{i\infty}$)
    and the composition of the forces exerted on the electrons and
    ions, as $v_h$ is varied.}
  \label{fig:forceplot}
\end{figure}
Fig.\ \ref{fig:forceplot} illustrates the search process
graphically. Panel (a) simply plots, in the fixed ion frame, the
distant ion velocity $f_{i\infty}(v_\infty)$ distribution used for the
examples we are giving. Panel (b) shows as a function of $v_h$ the
total force on ions, electrons, and their sum, when $\Delta\phi$ is
such that $n_e(\pm\infty)=n_i(\pm\infty)$, that is, distant neutrality
is satisfied. Notice that there is substantial cancellation between
$F_i$ and $F_e$. The total force $F_i+F_e$ given in blue is the
critical quantity. Equilibria occur where it is zero. This scan shows
that there are three such $v_h$ roots. However, at two of them, the
ones located near the distribution maxima, the slope $dF/dv_h$ is
positive. That sign means that at any adjacent velocity the non-zero
force acts to accelerate the hole potential velocity $v_h$ \emph{away}
from the equilibrium value. Thus those equilibria are unstable to slow
acceleration. Therefore the $F$ zero that is of interest is the middle
one where $dF/dv_h<0$, and is selected by the scan for further
refinement of the $v_h$ value. The vertical line and the cross on the
$f_{i\infty}$ plot indicate that equilibrium value.

\begin{figure}[htp]
  \centering
  \includegraphics[width=0.7\hsize]{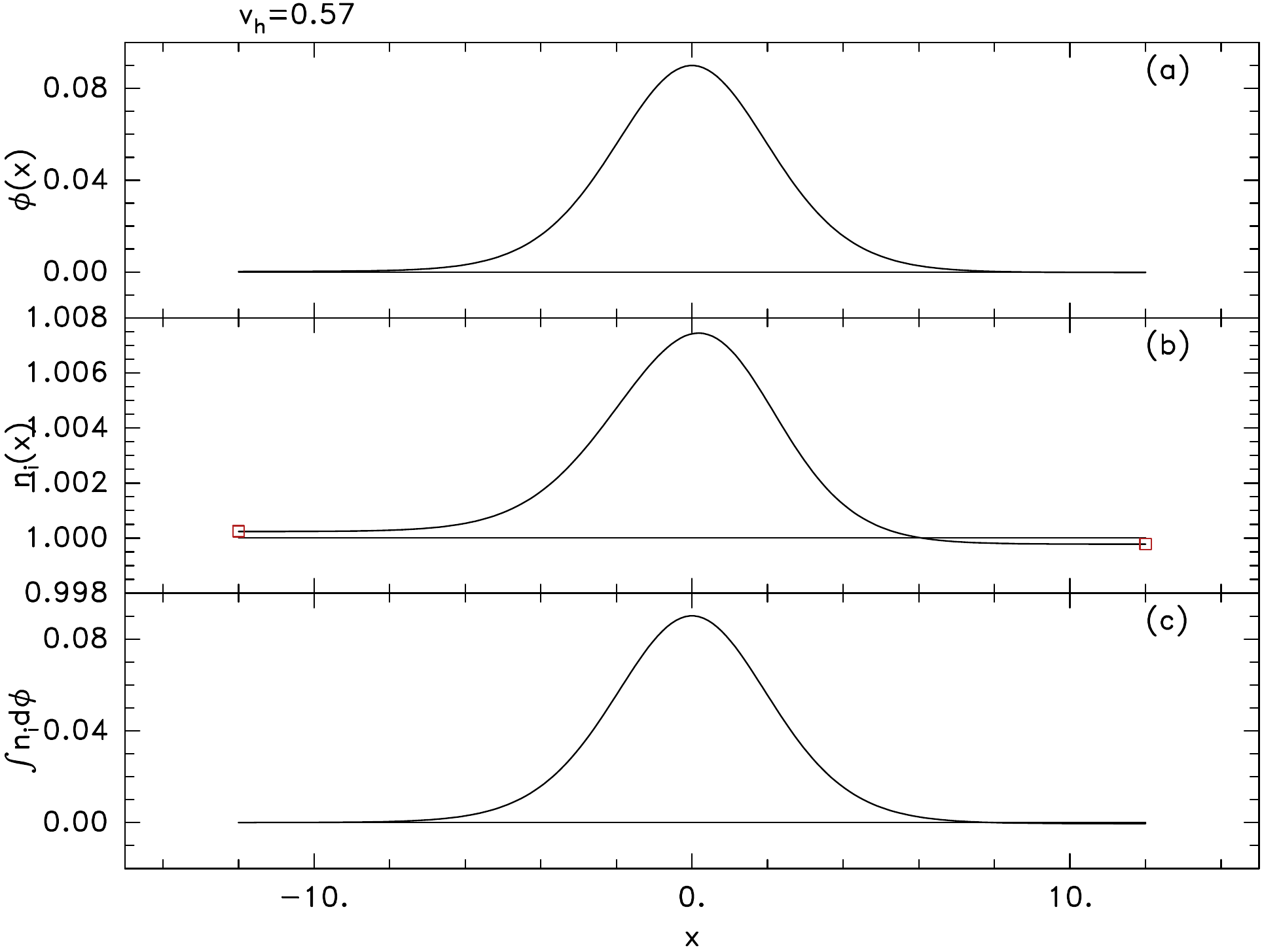}
  \caption{Converged equilibrium shape of potential $\phi(x)$, ion
    density $n_i(x)$ and ion force, when the hole required velocity
    $v_h=0.57$ for equilibrium has been discovered. For the same
    $\psi$ and $f_{i\infty}$ as before.}
  \label{fig:convplot}
\end{figure}
Fig.\ \ref{fig:convplot} shows the resulting potential, density and
force distributions as a function of position when the hole speed
corresponds to equilibrium. Now the total force is zero
$\int_{-\infty}^\infty (n_i-n_e){d\phi\over dx} dx=0$, and to achieve this the shift of
the hole-frame ion distribution (by $-v_h$) has almost (but not quite)
symmetrized the densities and potentials:
$n_i(+\infty)\simeq n_i(-\infty)$,
$\phi(+\infty)\simeq \phi(-\infty)$. The remaining asymmetry of the
distant ion density is cancelled by a very small potential difference
changing the electron density. It should be emphasized that this near
spatial symmetry only occurs at or near equilibrium. At $v_h$ values
in Fig.\ \ref{fig:forceplot} where the forces are large, there are
much larger asymmetries in the three curves than Fig.\ \ref{fig:convplot}
shows; compare Fig.\ \ref{fig:explot}, where $v_h=0$, for example.

\subsection{Potential Shape}

The shape $\phi(x)$ is so far undetermined except for the values of
its extrema. We are therefore in the usual situation for BGK modes of
having great liberty in the potential shape, depending on the velocity
distribution of trapped electrons. (The $\phi(x)$ plots given so far
should be considered illustrative of plausible possibilities.) The
difference in potential $\Delta \phi$ is of course, induced by
ion-density differences. So ions must be accounted for in relating the
trapped electron distribution to $\phi(x)$. The simpler choice is to
regard $\phi(x)$, rather than $f_{et}(v)$, as prescribed in the
trapping region, and deduce the required trapped electron distribution
by solving the integral equation that arises from setting
$e(n_i-n_e)=-\epsilon_0{d^2\phi\over dx^2}$. Naturally, there will be
some constraints on the trapped distribution such as non-negativity
and finite slope. But these should be no more difficult to satisfy
than they are for symmetric holes. Moreover, it is not actually
necessary to solve to find $f_{et}(v)$ in order to complete the hole
structure determination, because it is only the trapped
density $n_{et}(\phi)$ that is required.

We are, however, not now free to choose separately the $\phi(x)$
profiles on \emph{both sides} of the hole ($\sigma_x=\pm1$), because
the trapped electron density at a particular potential is the same on
both sides. The ion density is not symmetric, but is already
prescribed on both sides. Let
$\sigma_\infty=\sigma_m\equiv sign(\Delta \phi)$ denote the side with
higher distant potential. Then there is no freedom to adjust the
potential profile by trapped electron distribution choices at energies
$-e\phi(\sigma_m\infty)<\energy_e<-e\phi(-\sigma_m\infty)$, because
those electrons are not trapped, they are reflected. If we freely
prescribe the potential profile for side $\sigma_x=\sigma_m$, it
determines the required trapped velocity distribution. But then the
spatial profile on the other side ($\sigma_x=-\sigma_m$) must be found
by solving Poisson's equation there, because the trapped electron
distribution has already been determined.  By virtue of the way we
chose $\Delta\phi$ and $v_h$ to satisfy equilibrium, all the boundary
conditions on the $-\sigma_m$ side can be satisfied; that is, if we
start the Poisson solution with $\phi=\psi$ and $d\phi/dx=0$ at $x=0$,
we will find that $\phi(\infty)=-|\Delta\phi|/2$ and
${d\phi\over dx}|_\infty=0$.  The natural way to solve for the entire
profile is to use the implicit form
$x=\pm\int_\psi^{\phi(x)} \sqrt{\epsilon_0/2|V|}d\phi$. This integration
has been implemented numerically, and gives results consistent with
the boundary conditions at infinity, to an accuracy dependent on the
fineness of the integration grids. 

The illustrative form of the potential prescribed on side $\sigma_m$
is chosen to be
\begin{equation}\label{phishape}
  \phi(x)=\phi_\infty+(\psi-\phi_\infty){\exp(L)+1\over
  \exp(L)+\cosh^4(x/4\lambda)},
\end{equation}
where the adjustable parameter $L$ when positive is the approximate
length of a flattened region at the top of the potential, and when
negative rapidly suppresses flattening; and $\lambda$ controls the
distant exponential decay, usually being the (generalized) Debye
screening length. This yields electron trapped velocity distributions
of approximately the (negative temperature $T_{et}$) Maxwellian form
$\propto\exp(-\energy/T_{et})$, when $L\to-\infty$\cite{Hutchinson2017}.
\begin{figure}[htp]
  \includegraphics[width=0.48\hsize]{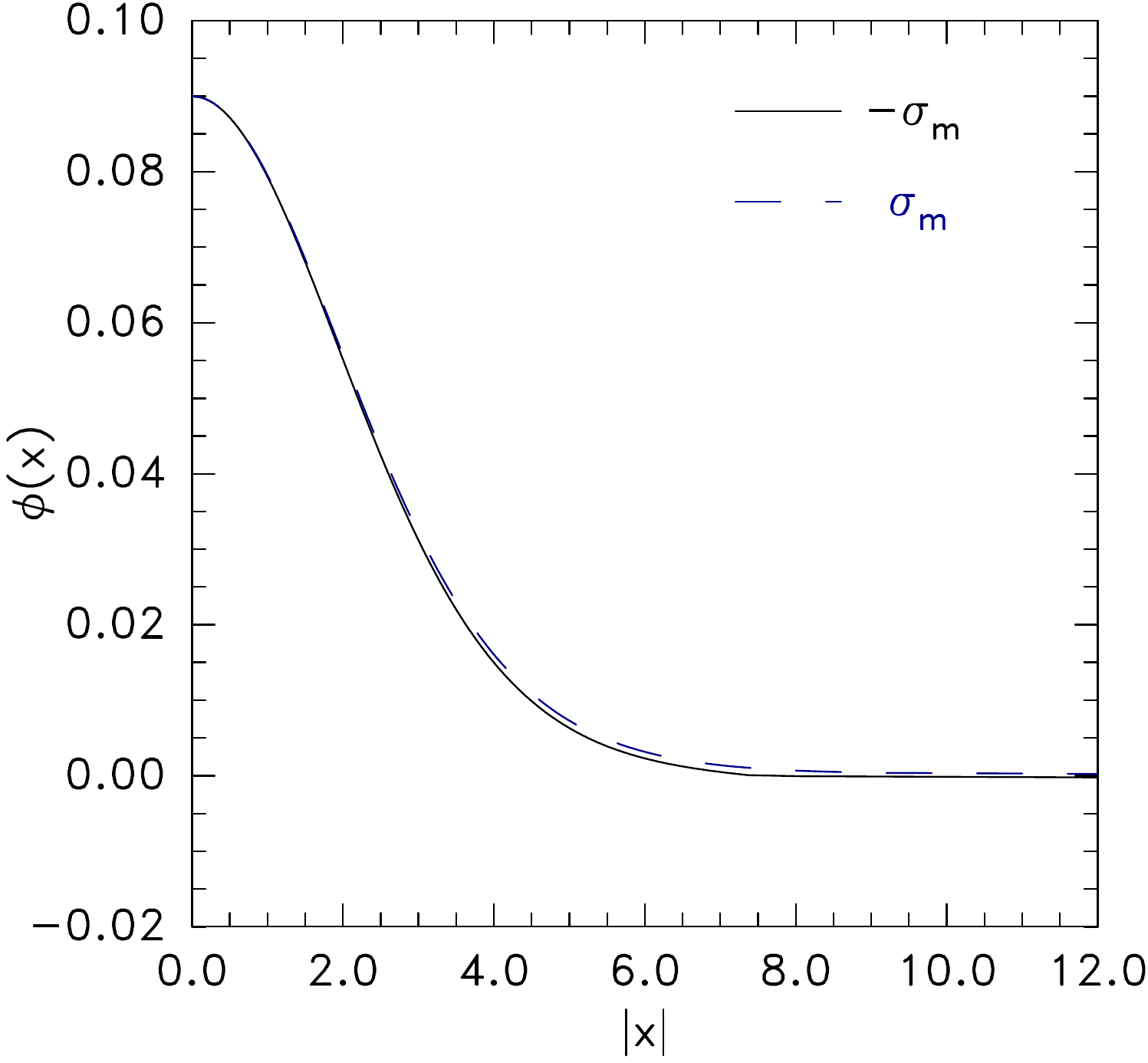}
  \includegraphics[width=0.48\hsize]{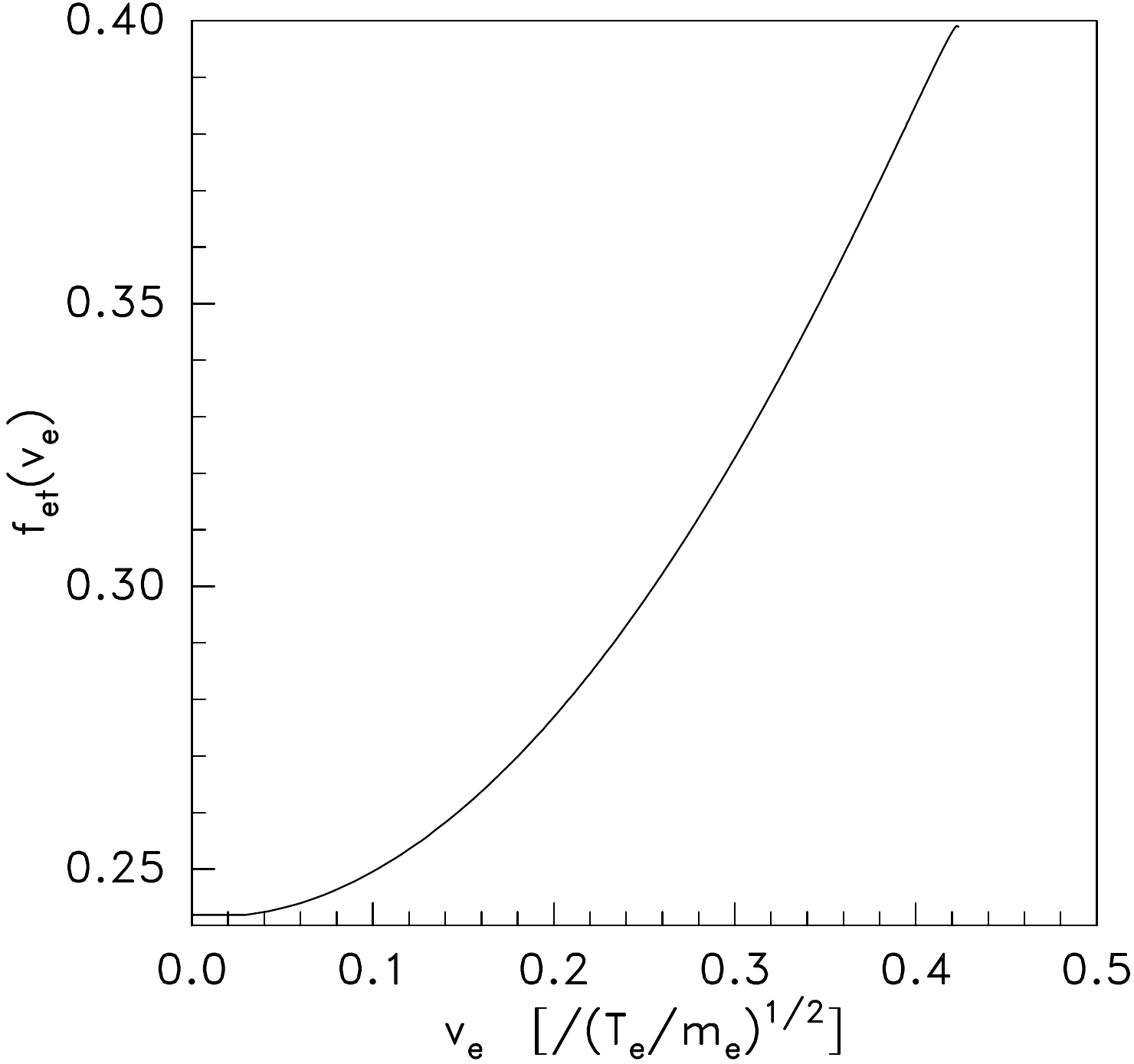}
  \caption{(a) Potentials on the two sides of the hole. Prescribed
    side dashed line, Poisson solution side solid. (b) Corresponding trapped
    electron distribution function at $x=0$, $\phi=\psi$.}
  \label{fig:phiofmodx}
\end{figure}
Fig.\ \ref{fig:phiofmodx}(a) shows (for $L=-10$, $\lambda=1$ and the
$f_{i\infty}$ of Fig.\ \ref{fig:forceplot}(a)) that the potential
shape derived from the solution of Poisson's equation (side
$-\sigma_m$, solid curve) is very close to that for the prescribed
side $\sigma_m$: eq.\ (\ref{phishape}). The trapped electron
distribution derived from solving the integral equation is shown in
Fig.\ \ref{fig:phiofmodx}(b).

\subsection{Stability to fast acceleration}

In addition to the steady-state force imbalance already discussed, an
additional mechanism that could give rise to hole velocity instability
involves force imbalance arising from hole acceleration itself. An
electron hole accelerating on the electron response timescale does not
permit adjustment of the ion density fast enough to be effectively
steady. As an approximation based on the separation of timescales, one
can approach this issue by supposing there is negligible ion density
change during some fast shift of the electron hole's velocity and
position from a full steady equilibrium. There would then arise a net
force change of the hole potential on the ions (but by assumption not
on the electrons) due to the shift displacement $\delta x$ of the
potential structure from the original equilibrium, acting on the
undisplaced ion density. For small displacements, the linearized
change in potential at any position is $-{d\phi\over dx} \delta x$,
giving rise to a change in force
$\delta F = - e\delta x \int -{d^2\phi\over dx^2} n_i(x) dx = -e\delta
x \int{d\phi\over dx} {dn_i\over dx} dx=-e\delta x \int {dn_i\over dx}
d\phi$. Notice that the integral is of two (approximately)
antisymmetric quantities ${d\phi\over dx}$ and ${dn_i\over dx}$, so it
is finite regardless of potential shape, but has the sign of
$-dn_i/d\phi$ of the dominant contributions to the integral. Stability
depends on the sign of $\delta F/\delta x$, and therefore of the
integral. If it is such as to enhance $\delta v_h$ and hence
$\delta x$, which arises if $\delta F/\delta x$ is positive,
exponential growth will occur. In so far as the system is correctly
described dynamically by this shift motion with static ions, it will
be stable if both $\delta F/\delta x$ and the equilibrium quantity
$dF/dv_h$ are negative. Given the equilibrium solution, it is easy
(numerically) to evaluate $\delta F/\delta x$, and it can be used to
qualify an equilibrium's dynamic as well as static stability. 

Since the potential asymmetry is generally very small, and the
$\delta F/\delta x$ depends only weakly on the $\phi(x)$ shape, it is
usually sufficient to calculate it approximately using a model $\phi$
profile whose Poisson-solution side is approximated as equal to the
specified-side's $\phi(x)$ matched at $\phi=\psi$ with
$\phi-\psi$ scaled to give the known $\Delta\phi$. That is what is
plotted in figures \ref{fig:explot} and \ref{fig:convplot}. Varying
the $x$-scale-length on either side makes no difference.

\begin{figure}[htp]
  \centering
  \includegraphics[width=.6\hsize]{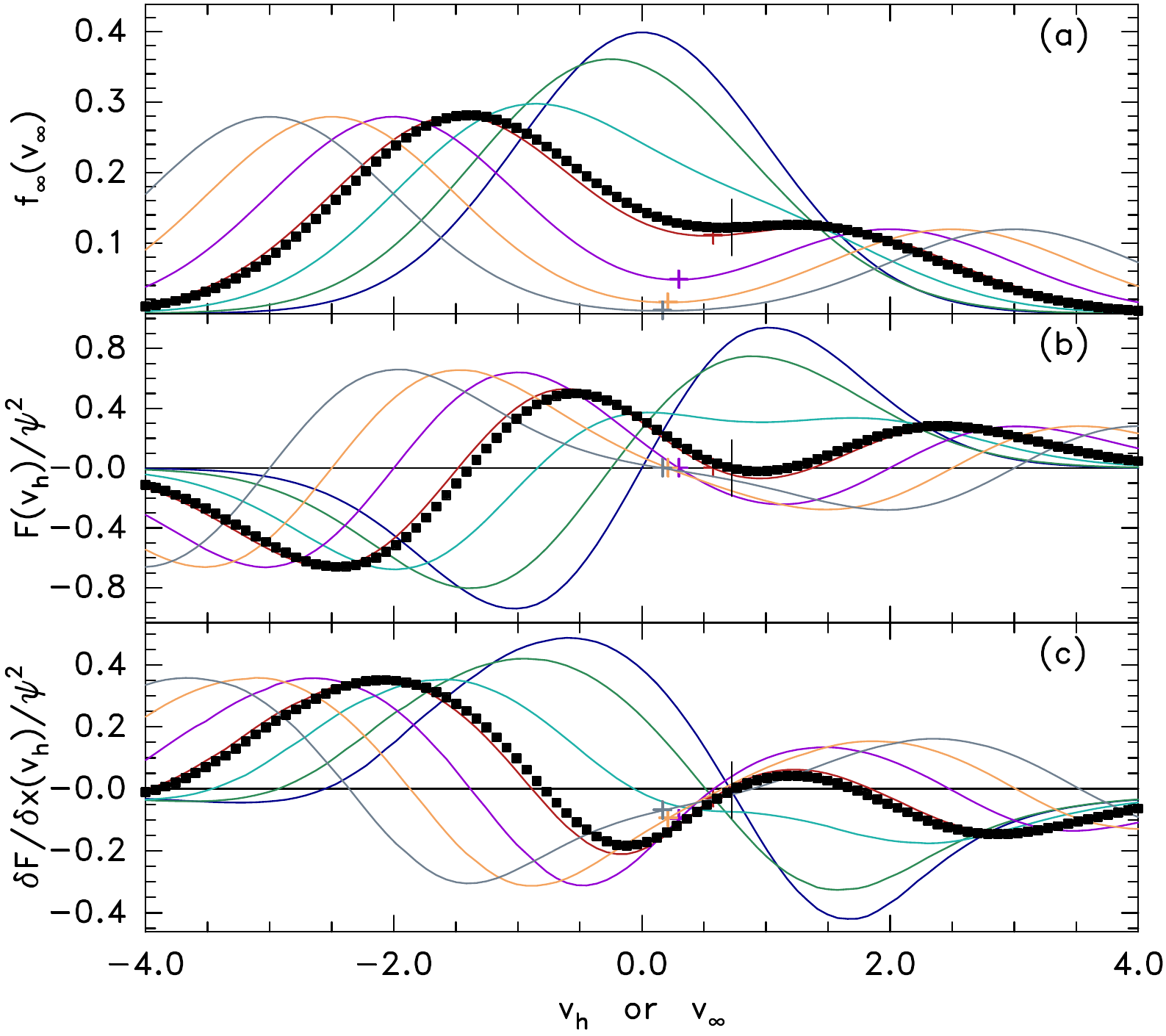}
  \caption{Stability parameters as a function of $v_h$, for a range of
    distributions obtained by scaling the Maxwellian components' shift
    velocity.}
  \label{fig:dynamic}
\end{figure}

It is valuable to explore the existence of a stable equilibrium for a
\emph{range} of ion distribution shapes. One way to do this is to
scale the velocity shift of each Maxwellian component (but not their
width or density) by a range of factors. The result of such a set of
calculations is shown in Fig.\ \ref{fig:dynamic}.
It includes (a) the velocity distribution, $f_{i\infty}(v_{\infty})$
(b) the steady force $F(v_h)$, and (c) the dynamic-shift force
coefficient $\delta F/\delta x$ for a set of 7 different velocity
shift scalings, colored by shift factor. The middle scaling factor is
1 (red) and corresponds to the distribution of the previous 5
figures. It confirms that $\delta F/\delta x$ is negative for the
equilibrium $v_h$ shown by the cross, as previously found. Therefore
by the approximate dynamic analysis this equilibrium is stable. For
zero shift factor (blue) the ion distribution is a single
Maxwellian. It's equilibrium $F(v_h)=0$ is unstable, and in such cases
no cross is plotted. For a large shift factor of 2 (grey) the
distribution consists of two components hardly overlapping. All
distributions plotted from scale-factor of $\sim1$ upward are
stable. All below are unstable. A refined intermediate value of the
scaling factor at the threshold for stability is found and plotted as
black points with the corresponding equilibrium $v_h$ indicated by a
short vertical line.  It is noticeable that the static $dF/dv<0$ and
dynamic $\delta F/\delta x$ stability thresholds are the same for this
coarse scan. In other words (approximately): if and only if a
distribution allows a statically stable equilibrium, it is dynamically
stable. Moreover, it is required to have a local minimum in
$f_{i\infty}$ in order for a stable equilibrium to exist. As has been
shown previously\cite{Hutchinson2021c}, the larger the $\psi$ the
deeper the minimum has to be. But for this moderate case,
$\psi=0.09 T_0$ (and smaller $\psi$), the depth required is small.

\subsection{Summary of Algorithm and Numerical Implementation}

The algorithm that has been implemented numerically is this.

\paragraph{1. From specified $f_{i\infty}(v_\infty)$} --- which is in
principle arbitrary but is conveniently represented by a sum of
shifted Maxwellian components of different densities and temperatures
--- and specified potential peak $\psi$, integrate equation
(\ref{ni}), using equation (\ref{eq:fiinf}) to obtain
$n_i(\phi,\sigma_x)$.

\paragraph{2. Determine the potential asymmetry $\Delta \phi$} using a
search followed by Newton iteration of $\Delta\phi$, so as to satisfy
equation (\ref{fedist}). The algorithm of step 1 is used to give each
iteration's $n_i(\phi(\pm\infty),\sigma_x)$.

\paragraph{3. Find the equilibrium hole velocity $v_h$} by searching,
with repetitive use of step 2 to evaluate the total force $F$ exerted
by the potential on particles. This is implemented by a coarse scan of
$v_h$ (which provides data for explanatory plots such as Fig.\
\ref{fig:dynamic}) followed by iterative refinement of the precision
of the equilibrium $v_h$ making $F=0$.

\paragraph{4. Construct the $\phi(x)$} by prescribing the higher
potential side's ($\sigma_m$) $\phi(x)$ using potential form eq.\
(\ref{phishape}), giving the trapped electron density
$n_{et}(\phi)$. Solve Poisson's equation on the other side
($-\sigma_m$) using the then known $n_i(\phi)$, $n_e(\phi)$. Verify
whether $\delta F/\delta x$ satisfies dynamic stability.  If desired,
solve the integral equation to find the trapped electron distribution
function from the prescribed $\phi(x)$.

\section{Asymmetric Holes at a range of parameters}

\begin{figure}[htp]
  \centering
  \includegraphics[width=.9\hsize]{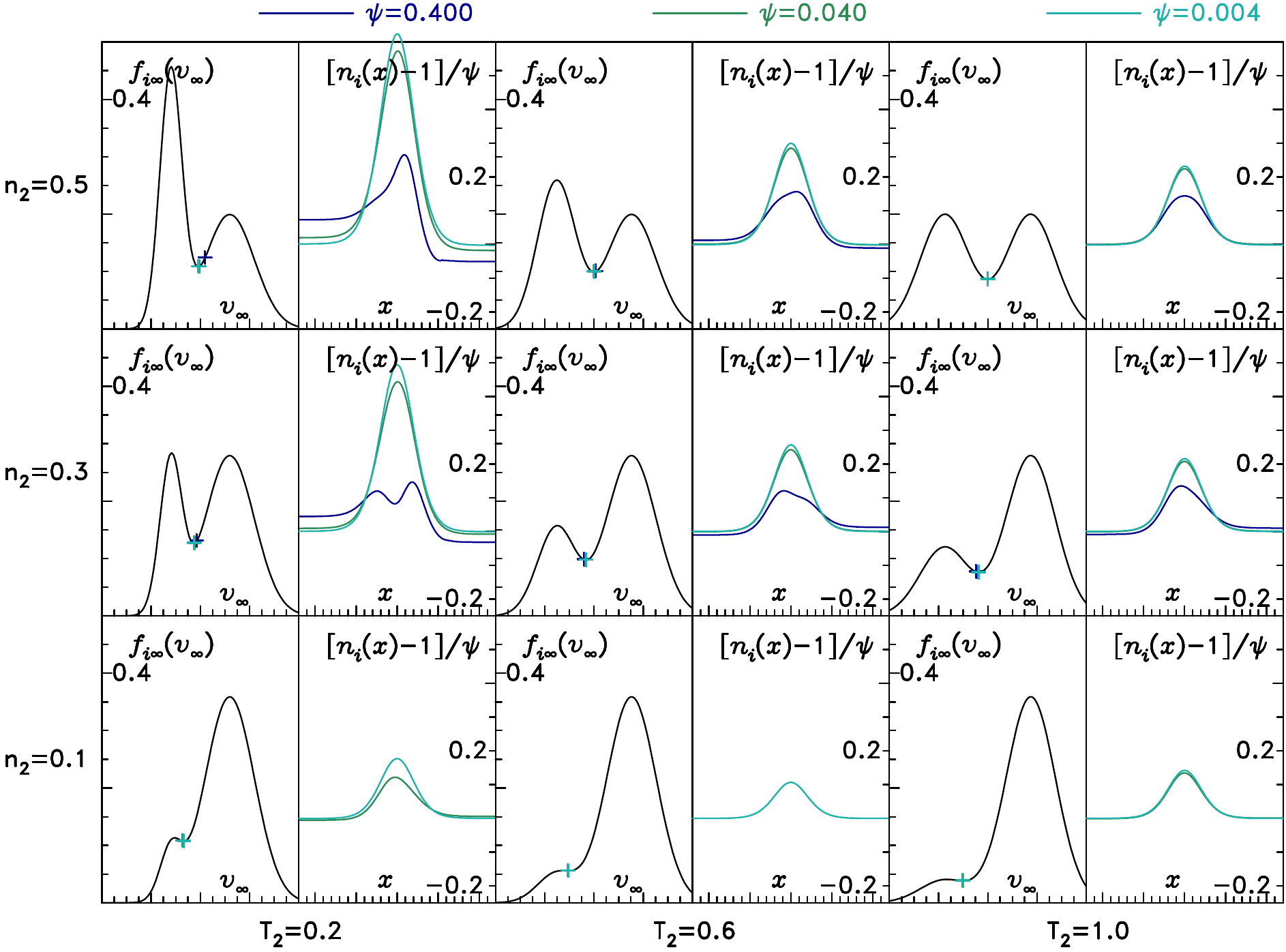}
  \caption{A 3$\times$3 display of 9 different cases
    $T_2$=$(.2,.6,1)$, $n_2$=$(.1,.3,.5)$ for which
    $v_s=0.75+\sqrt{T_2}$; here $T_e=T_0$. Each case has two adjacent
    subframe plots: the left-hand is $f_{i\infty}(v)$ and the
    right-hand represents scaled ion density perturbation as
    $[n_i(x)-1]/\psi$. All corresponding frames have the same axis
    ranges $-4\le v_\infty\le 4$, $-12\le x \le 12$. Densities are
    plotted for 3 different potential peak heights
    $\psi$=$(.4,.04,.004)$ indicated by different colors, but only if
    a stable equilibrium hole velocity ($v_h$) exists. If it does, a
    cross is plotted of that color on the left-hand subframe at that
    velocity and distribution height.}
  \label{fig:multiden}
\end{figure}

We now illustrate asymmetric hole equilibria with $f_{i\infty}$
consisting of two shifted Maxwellian components, having densities
$n_{1,2}$, temperatures $T_{1,2}$, and velocities in the reference
(not hole) frame $v_{1,2}=\pm v_s$. The reference temperature is taken
equal to the temperature of component one $T_0=T_1$, and the sum of
the densities (the total background ion density) is unity
$n_1+n_2=1$. Thus $n_2$, $T_2$, and $v_s$ are the three parameters
determining the distribution shape. Fig.\ \ref{fig:multiden} surveys
the shapes $f_{i\infty}(v)$ and resulting ion density profiles
$n_i(x)$.

The overall scaling of the ion density perturbation is seen to be
$n_i-1\sim \psi$. This is intuitive; the amplitudes of the potential
perturbation and the density perturbation are proportional.  We
observe that there is very little difference in the scaled density
perturbation between the lower amplitude cases $\psi=0.04,0.004$, even
quantitatively. This is because for these cases the deficit in
$f_e(v)$, and of $n_i$, supporting the hole, is fractionally small and
can be linearized. By contrast, for $\psi=0.4$, the hole is deep and the
ion density response is no longer linear. Moreover, for the bottom
row, none of the distributions permits a stable hole at $\psi=0.4$
even though it does at the lower amplitudes. That is because deeper
holes would require a deeper local minimum in $f_{i\infty}$ for
stability than is provided by these cases.

The top right case ($T_{2}=1$, $n_2=0.5$) is completely symmetric, and
the equilibrium is at $v_h=0$, the symmetry axis. Also there is no ion
density difference across the hole and the corresponding potential
difference $\Delta\phi=(T_e/e)\{\ln[n(+\infty)]-\ln[n(-\infty)]\}$ is
zero. At lower values of $T_2$, substantial asymmetry in $n_i$
appears, becoming quite pronounced for $T_2=0.2$, (at the left)
regardless of $n_2$ for the largest $\psi$. For smaller $\psi$,
density $n_i(\infty)$ becomes more symmetric, although some small
asymmetry remains, most noticeable at low $T_2$. Overall, though, the
asymmetry in $n_i(\infty)$ at equilibrium remains less than
$\sim 0.2\psi$, and in fact scales like $\psi^2$.

\section{Algebraic Calculation of Asymmetry}

\subsection{Explanation of approach}
The scaling of potential asymmetry can be understood qualitatively as
follows. We recognize that there are two types of asymmetric
contribution to the neutrality requirement ($\Delta n_e=\Delta n_i$)
and the force balance requirement ($F=0$). There is an
\emph{intrinisic} asymmetry arising even when $\Delta\phi=0$ that
comes from the asymmetry of $f_{i\infty}$, and there is an
\emph{extrinsic} asymmetry that comes from $\Delta \phi$.  The
asymmetry (strictly antisymmetric part) of $f_{i\infty}$ can be
expressed as the odd terms of its Taylor expansion in velocity in the
hole frame $f'v+{1\over6}f'''v^3+O(v^5)$, $f'$ and $f'''$ are the
first and third velocity-derivatives of $f_{i\infty}$ evaluated at
$v=0$ (i.e.\ at ion-frame velocity $v_h$) taken as constants.

The intrinsic ion density asymmetry, which is to be evaluated between
the positions on either side of the hole corresponding to
$\phi=+|e\Delta \phi|/2$, has then two terms proportional respectively
to $f'\psi$ and $f'''\psi^2$. The $\psi$ dependences arise from
integrals over $v$ to the hole height energy
$v=\sqrt{2e\psi/m}$. There is no intrinsic electron density asymmetry
because electrons are trapped by the positive potential peak, not
reflected from it. The extrinsic density asymmetry arises from the
change of $n_e$ and $n_i$, on the low potential side, across the range
$-|\Delta\phi|/2 \le \phi \le |\Delta\phi|/2$. These changes are both
proportional to $\Delta\phi$, but with different coefficients.

The intrinsic ion force asymmetry likewise has two terms
$\propto f'\psi^2$ and $\propto f'''\psi^3$, where the additional
power of $\psi$ (relative to $\Delta n$) comes from the integral
$\int n d\phi$. There is no intrinsic electron force. The total
extrinsic force arises from reflection of ions and electrons from the
potential range $-|\Delta\phi|/2 \le \phi \le |\Delta\phi|/2$, but it
is most easily expressed as the difference in the Maxwell stress on
the low potential side between $-|\Delta\phi|/2$ and $|\Delta\phi|/2$,
which is simply $(\epsilon_0/2\lambda^2)(\Delta\phi)^2$, where
$\lambda$ is the length for generalized Debye screening including the
response of both electrons and ions. Therefore the simplified structure of the
simultaneous equilibrium requirements (writing coefficients
$a,b,c,d$ to be found later) is
\begin{equation}
  \label{eq:simul}
  \begin{split}
    \Delta\phi &= a \psi f'+ b f'''\psi^2\\
    \Delta\phi^2 &= c \psi^2f'+ d f'''\psi^3.
    \end{split}
\end{equation}
Eliminating the $f'$ terms,
\begin{equation}
  \label{eq:Delphieq}
  \Delta\phi^2-{c\over a}\psi\Delta\phi-
  \left(d- {bc\over a}\right) f''' \psi^3=0,
\end{equation}
which can be solved for $\Delta\phi$ as
\begin{equation}
  \label{eq:Dpsolve}
  \Delta\phi={c\psi\over 2 a}\left[1\pm\sqrt{1+ \left(d
    - {bc\over a}\right)\left(2a\over c\right)^2f'''\psi}\;\right]
  .
\end{equation}
The discriminant's sign must be chosen to be opposite the sign of the
first term.  When $\Delta\phi/\psi$ is small, only the linear term is
important in eq.\ \ref{eq:Delphieq}, and
\begin{equation}
  \label{eq:Dpapprox}
  \Delta\phi
\simeq
  \left(b-{ad\over c}\right)f'''\psi^2.
\end{equation}
Thus both $\Delta\phi$ and $\Delta n_i$ scale $\sim \psi^2$, and also
$\sim f'''$, which is the lowest order contribution to asymmetry in
$f_{i\infty}$ about the local minimum where $f'=0$. Actually the
solution is not exactly at $f'=0$, but it is at a value that makes the
two equations consistent, which is
\begin{equation}
  \label{eq:fprime}
  f'=(\Delta\phi - bf'''\psi^2)/a\psi \simeq
 -(d/c) f'''\psi.
\end{equation}
And the distribution derivatives $f'$ and $f'''$ must be evaluated at
the $v_h$ value that satisfies this equation.

\subsection{Intrinsic ion density asymmetry}

To quantify our analytic asymmetry expressions we need to evaluate the
$a,b,c,d$ coefficients used in eqs.\ (\ref{eq:simul}) to
(\ref{eq:fprime}), to relevant order in $\psi$ and $\Delta\phi/\psi$.
Those equations have already assumed that $\psi$ is a small quantity
to allow the $f_{i\infty}$ expansion; eq.\ \ref{eq:Dpapprox} shows
that $\Delta\phi/\psi=O(\psi)$; and eq.\ \ref{eq:fprime} shows that
$f'/f'''= O(\psi)$. In this section for brevity we introduce a
notation for the equivalent velocity,
$v_{\psi-\phi}=\sqrt{2e(\psi-\phi)/m}$, and in a similar way
$v_\phi=\sqrt{2e\phi/m}$, $v_\psi=\sqrt{2e\psi/m}$, and
$v_\energy=\sqrt{2\energy/m}=\sqrt{v^2+v_\phi^2}$. We also work in
scaled units so that $e=1$, $m=1$, and $T_0=1$.

The intrinsic ion density asymmetry between two points on opposite
sides of the potential peak, at the same potential $\phi$, is found
from eq.\ (\ref{eq:fiinf}). Noting that only for reflecting energies
is there any asymmetry, we get
\begin{equation}
  \label{eq:Dnf}
  \Delta n_f(\phi)=2\int_0^{v_{\psi-\phi}} f_{i\infty}(-\sqrt{v^2+v_\phi^2})-
  f_{i\infty}(\sqrt{v^2+v_\phi^2})\,dv = 2\int_0^{v_{\psi-\phi}} \Delta f_i
  dv.
\end{equation}
 Substituting
the expansion for the antisymmetric part of $f_{i\infty}$, so that
$\Delta f_i= -(2f' v_\energy+{1\over 3}f'''v_\energy^3)$ we have
\begin{equation}
  \label{eq:nfexp}
  \Delta n(\phi)=-2\int_0^{v_{\psi-\phi}}[2f'(v^2+v_\phi^2)^{1/2}+{1\over 3}
  f'''(v^2+v_\phi^2)^{3/2}]dv.
\end{equation}
Performing the integrals, we find
\begin{equation}
  \label{eq:Dnint}
  \begin{split}
  \Delta n(\phi) =& -2[v_{\psi-\phi} v_\psi +v_\phi^2\ln({v_{\psi-\phi}+v_\psi\over v_\phi})]f'\\
  &-{1\over 12}[v_{\psi-\phi} v_\psi(2v_\psi^2+3v_\phi^2)+3v_\phi^4 \ln({v_{\psi-\phi}+v_\psi\over v_\phi})]f'''.
  \end{split}
\end{equation}
The lowest potential at which this applies is $\phi=|\Delta\phi|/2$,
because on the higher side there are no lower potentials. Then
$\Delta n_f=\Delta n(|\Delta\phi|/2)$ is the intrinsic contribution to the
neutrality criterion. But because of the smallness of $\Delta\phi/\psi$, to
lowest order we find a result independent of $\Delta\phi$.
\begin{equation}
  \label{eq:Dnint2}
  \begin{split}
  \Delta n_f&=-2v_\psi^2f'-{1\over 6} v_\psi^4f'''
  +O(v_\psi^2v_{|\Delta\phi/2|}^2)\\
  &=-4\psi f' -{2\over 3}\psi^2 f''' +O(\psi\phi)
  \end{split}
\end{equation}

\subsection{Intrinsic Ion Force}

The integral of eq.\ (\ref{eq:Dnint}) also provides us with the
intrinsic force exerted by the potential on ions,
$F_f=\int_{|\Delta\phi|/2}^\psi \Delta n(\phi) d\phi = {1\over
  2}\int_{v_{|\Delta\phi/2|}^2}^{v_\psi^2} \Delta n\;
dv_\phi^2$. Closed form indefinite integrals exist:
\begin{equation}
  \label{eq:int1}
  \int v_{\psi-\phi}v_\psi + v_\phi^2\ln({v_{\psi-\phi}+v_\psi\over v_\phi})dv_\phi^2
  ={1\over 2}[ v_{\psi-\phi}v_\psi(-2v_\psi^2+v_\phi^2)+v_\phi^4\ln({v_{\psi-\phi}+v_\psi\over v_\phi})]
\end{equation}
and
\begin{equation}
  \label{eq:int2}\begin{split}
    &\int [v_{\psi-\phi} v_\psi(2v_\psi^2+3v_\phi^2)+3v_\phi^4
    \ln({v_{\psi-\phi}+v_\psi\over v_\phi})]dv_\phi^2
    =\\
    &\qquad\qquad\qquad{1\over3}
    v_{\psi-\phi}v_\psi(-8v_\psi^4+2v_\psi^2 v_\phi^2+3v_\phi^4)
    +v_\phi^6\ln({v_{\psi-\phi}+v_\psi\over v_\phi}).
  \end{split}
\end{equation}
Noting that the upper limits do not contribute because at
$v_\phi=v_\psi$, $v_{\psi-\phi}=0$, the required definite integral
becomes, to lowest order\footnote{The structure of the second (force)
  equation of (\ref{eq:simul}) puts two terms that are intrinsically
  of order $\psi^3$ equal to a left hand side of order $\phi^2$ which
  at equilibrium is of order $\psi^4$. Therefore it might seem that we
  must keep terms up to $O(v_\psi^4v_{|\Delta\phi/2|}^2)=O(v_\psi^8)$;
  that is, the terms containing a single factor $v_\phi^2$. However,
  the elimination process with the first equation of (\ref{eq:simul})
  shows those terms are of order $\psi$ smaller in calculating
  $\Delta\phi$, and they can safely be ignored, as has been verified by
  numerical evaluation including the higher order terms.}
\begin{equation}
  \label{eq:Fintr}
  \begin{split}
    \iffalse
    F_f&=(-v_\psi^4+ {1\over 2}v_\psi^2 v_{|\Delta\phi/2|}^2)f' +
     {1\over
      9}(-v_\psi^6+{5\over 4}v_\psi^4 v_{|\Delta\phi/2|}^2 )f'''
    +O(v_\psi^2v_{|\Delta\phi/2|}^4)\\
    &=(-4\psi^2+\psi|\Delta\phi|)f'+{1\over9}(-8\psi^3+5\psi^2|\Delta\phi|)f'''
    +O(\psi|\Delta\phi|^2).
    \else
    F_f&=-v_\psi^4f' +
     -{1\over
      9}v_\psi^6f'''
    +O(v_\psi^4v_{|\Delta\phi/2|}^2)\\
    &=-4\psi^2f'+-{1\over9}8\psi^3f'''
    +O(\psi^2|\Delta\phi|).
    \fi
    \end{split}
\end{equation}

\subsection{Extrinsic density difference and force}

The neutrality condition requires in addition the change in $n_i-n_e$,
which occurs between potentials $-|\Delta\phi|/2$ and $|\Delta\phi|/2$
on the lower $\phi_\infty$ side ($-\sigma_m$). It can be written
\begin{equation}
  \label{eq:Dpext}
  \Delta n_{\Delta\phi}= ({dn_i\over d\phi} - {dn_e\over
    d\phi})\Delta\phi
  =-{\epsilon_0\over e}{\Delta \phi\over\lambda^2}.
\end{equation}
The electron density change, since electrons are Maxwellian, is simply
the Boltzmann factor, which gives ${dn_e\over d\phi}=n_{e0}e/T_e$. If
this were the only source of charge, then Poisson's equation would be
${d^2\phi\over dx^2}=\phi/\lambda_{De}^2$, where
$1/\lambda_{De}^2=n_{e0}e^2/\epsilon_0T_e$ which in (the Debye)
normalized units is $T_0/T_e$.  Thus
${dn_e\over d\phi}e/\epsilon_0=1/\lambda_{De}^2=T_0/T_e$ gives the
shielding length due to electrons alone; and we can write
$-({dn_i\over d\phi} - {dn_e\over
  d\phi})e/\epsilon_0=1/\lambda^2=T_0/T_s=1/T_s$ (normalized)
expressing the modified shielding length $\lambda$ including the
linearized dielectric response of both electrons and ions in terms of
an effective shielding temperature $T_s$ which is generally
$\simeq T_e$.

Similarly, as previously noted, the extrinsic force on the particles
can be expressed using $\lambda$ as the difference in the Maxwell
stress in the range $-|\Delta\phi|/2$ ($|x|=\infty$) and
$|\Delta\phi|/2$, in which the potential is exponential with decay
length $\lambda$. Thus
$|F_{\Delta\phi}|= {\epsilon_0\over2}\left(d\phi\over dx\right)^2=
{\epsilon_0\over 2\lambda^2}\Delta\phi^2$, with sign $\sigma_\Delta$
equal to that of $\Delta\phi$:
\begin{equation}
  \label{eq:forceco}
F_{\Delta\phi}=sign(\Delta\phi){1\over 2T_s}\Delta\phi^2\equiv\sigma_\Delta{1\over 2T_s}\Delta\phi^2.  
\end{equation}

\subsection{Comparison with numerics}

Dividing the neutrality $\Delta n_{\Delta\phi}+ \Delta n_f=0$ and
force balance $F_{\Delta\phi}+F_f=0$ conditions by minus the
coefficients of their extrinsic terms $-1/T_s$ and
$\sigma_\Delta/2T_s$, we obtain the coefficients for eq.\
(\ref{eq:simul}):
\begin{equation}
  \label{eq:coefs}
  \begin{array}{cc}
    a=4T_s&b={2\over 3}T_s\\
    c=-8\sigma_\Delta T_s& d=-{16\over 9}\sigma_\Delta T_s
  \end{array}.
\end{equation}
Substituting for them we find the potential asymmetry
\begin{equation}
  \label{eq:subst}
  \Delta\phi=\sigma_\Delta\psi(1 -\sqrt{1 -\sigma_\Delta (4T_s/9)f'''\psi})
  \simeq {T_s2\over9}f''' \psi^2
\end{equation}
and $f'\simeq -{2\over 9}f'''\psi$. The density asymmetry is $\Delta
n_i=\Delta n_e\simeq \Delta\phi/T_e = T_s/T_e {2\over 9}f'''\psi^2$.

\begin{figure}[htp]
  \centering
  \includegraphics[width=0.42\hsize]{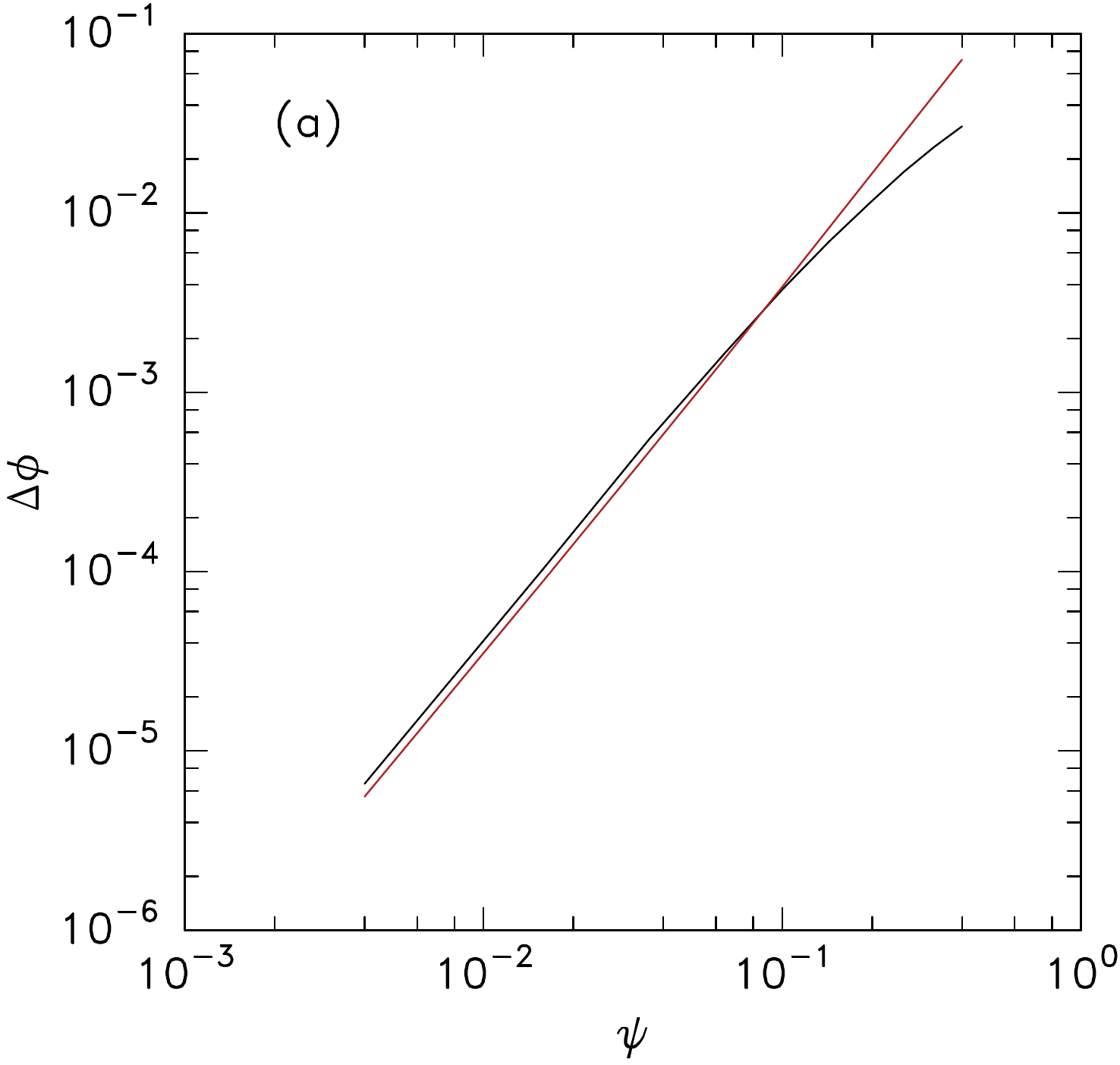}
  \includegraphics[width=0.5\hsize]{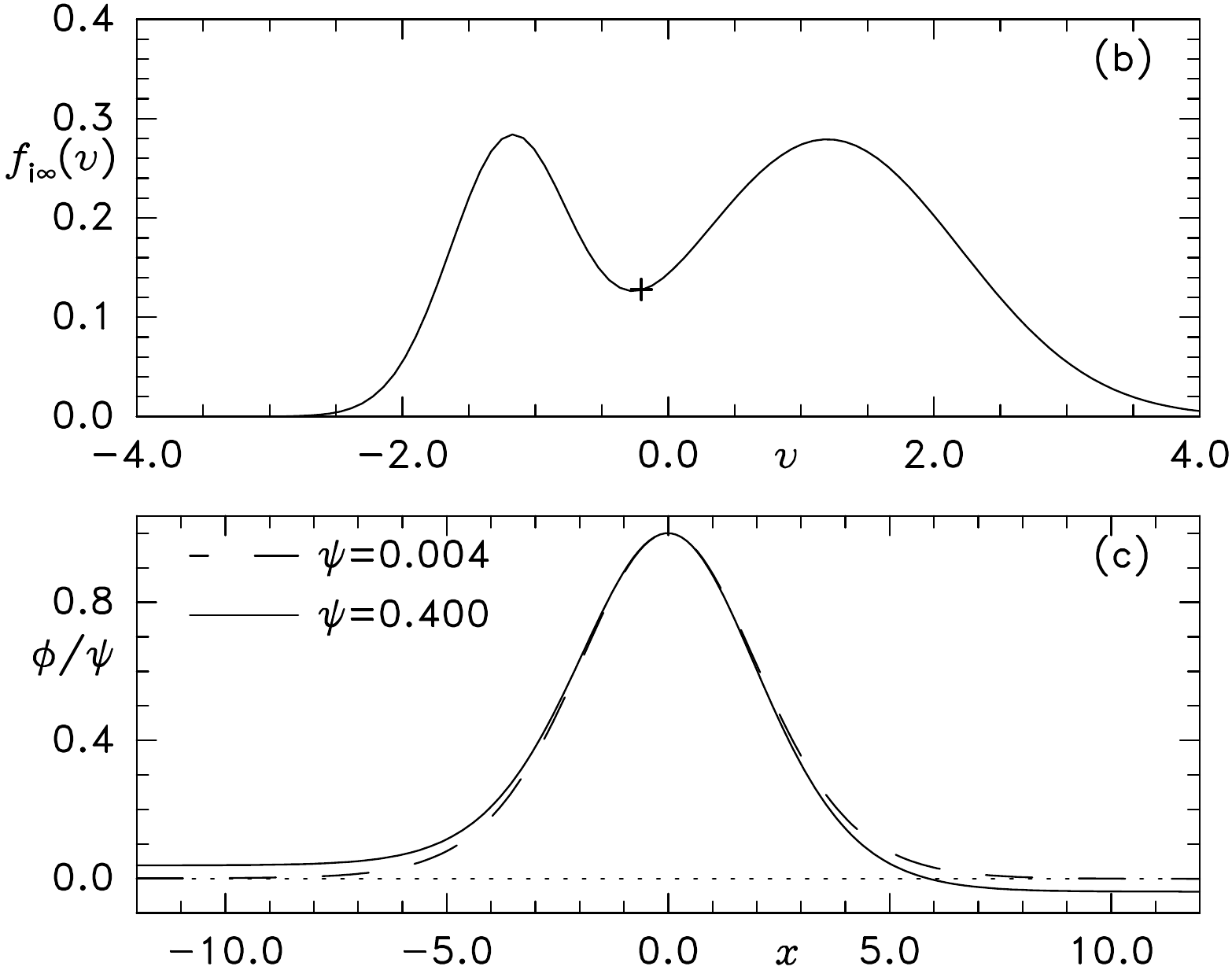}
  \caption{(a) Comparison of numerical (black) and algebraic (red)
    estimates of potential asymmetry $\Delta\phi$, showing its scaling
    proportional to the square of the electron hole peak potential
    $\psi$. In (b) is shown the ion distribution comprising Maxwellians $T_1=1$,
    $n_1=0.7$, $v_1=1.197$ $T_2=0.2$, $n_2=0.3$, $v_2=-1.197$, and
    $T_e=1$, in scaled units. The potential profiles in (c) are from the ends
    of the scaling range.}
  \label{fig:scaled}
\end{figure}

The ion response ${dn_i\over d\phi}$ contribution to
$T_s= ({1\over T_e}-{dn_i\over d\phi})^{-1}$ is estimated on an ad hoc
basis to be typically $0.3$. This estimate is probably the biggest fractional
uncertainty unless $T_e$ is very small (in which case the ion
contribution to shielding is small). Then we compare the present
algebraic estimate with the numerical evaluation in
Fig. \ref{fig:scaled}(a) .
The good agreement between the two except in the region where $\psi$
is no longer small, is gratifying and serves as a verification of the
numerical and algebraic integrations. The potential profiles
Fig. \ref{fig:scaled}(a) emphasize that even at the upper end of the
scaling range (where the agreement is compromised by $\psi$ no longer
being small) the asymmetry in the potential at equilibrium remains
small. This case, corresponds to the $n_2=0.3$, $T_2=0.2$ case of
Fig.\ \ref{fig:multiden}, correcting any false impression given there
by the density profiles that the potential asymmetry is strong. It never is.

\section{Discussion}

This numerical and algebraic analysis of asymmetric electron holes is,
to my knowledge, the first that treats plausibly realistic external
ion distributions taking into account the criteria of equilibrium. It
shows that for truly solitary equilibrium positive potential
structures, sustained by the plasma velocity distributions not imposed by local
constraints, potential asymmetry is only minor, and for small
amplitude electron holes is negligible.  This finding moderates past
speculations about asymmetric electron holes in the Double Layer
literature. It also justifies and confirms the recent
theory\cite{Hutchinson2021c} of slow electron holes that ignores
potential asymmetry. The present treatment remains purely an
equilibrium theory, but the force on the hole out of equilibrium has
been calculated, and its sign determines whether or not the equilibrium
is stable. For a stable equilibrium to exist, the ion distribution
must have a local minimum, and the hole velocity must lie within it.

It is possible for an electron hole to be \emph{formed} at a velocity
that does not satisfy the equilibrium force constraint, or in an ion
distribution shape that causes any slow equilibrium to be unstable. If
so, then it might initally have substantial potential asymmetry, or
develop it during unstable acceleration. But once a hole finds a stable
equilibrium velocity, that asymmetry will be largely suppressed. If,
therefore, a substantially asymmetric electron hole were to be
convincingly observed, its asymmetry might be an indication that it
was young, dynamic, and still in the process of accelerating toward
equilibrium. Of course, this treatment is also only one-dimensional,
and all of its conclusions should be qualified by the possibility of
being changed by multidimensional effects.

\section*{Acknowledgments}
I am grateful for discussions and collaboration with Ivan Vasko and
his colleagues about observations of electron holes in space. No
external funding supported the present work\footnote{The code that
  produced the figures in this paper is available at
  \url{https://github.com/ihutch/asymhill}}.

\bibliography{JabRef}

\end{document}